\documentclass[twocolumn,showpacs,preprintnumbers,superscriptaddress,amsmath,%
amssymb,nofootinbib]{revtex4}

\input psfig.sty

\usepackage{graphicx}
\usepackage{dcolumn}
\usepackage{bm}

\newcommand{\be}{\begin{equation}}
\newcommand{\ee}{\end{equation}}

\begin{document}

\title{Coupled quintessence with a possible transient accelerating phase}

\author{F. E. M. Costa} \email{ernandes@on.br}

\affiliation{Observat\'orio Nacional, 20921-400, Rio de Janeiro -- RJ, Brasil}

\date{\today}

\begin{abstract}

We discuss some cosmological consequences of a general model of coupled quintessence in which the phenomenological coupling between the cold dark matter and dark energy is a function of the cosmic scale factor $\epsilon(a)$. This class of models presents cosmological solutions in which the Universe is  currently dominated by an exotic component, but will eventually be dominated by cold dark matter in the future. This dynamical behavior is considerably different from the standard $\Lambda$CDM evolution, and may alleviate some conflicts in reconciling the idea of the dark energy-dominated universe with observables in String/M-theory. Finally, we investigate some observational features of this model and discuss some constraints on its parameters from current SNe Ia, BAO and CMB data.

\end{abstract}


\maketitle

\section{Introduction}

One of the main open problems in Cosmology is to determine the physical mechanism behind the current cosmic acceleration. This phenomenon has been evidenced by a combination of observational data~\cite{data} and, in the context of the general relativity theory, can be explained only if we admit the existence of an exotic field, the so-called dark energy. The origin and nature of this  exotic component constitute a complete mystery and represents one of the major challenges not only to cosmology but also to our understanding of fundamental physics~(see, e.g., \cite{review} for more on this subject). By assuming a spatially flat geometry, this mysterious component accounts for (in units of the critical density) $\simeq 0.7$ of the cosmic composition, a value that is of the same order of magnitude of the relative density of the cold dark matter, $\simeq 0.3$. However, since these dark components are usually assumed to be independent and, therefore, scale in different ways, this would require an unbelievable coincidence, the so-called coincidence problem (CP). 

A phenomenological attempt at alleviating the CP is allowing the dark matter and dark energy to interact. This phenomenology in turn gave origin to the so-called models of coupled quintessence, which have been largely explored in the literature \cite{cq, jesus, ernandes, ernandes1}. These scenarios are based on the premise that, unless some special and unknown symmetry in nature prevents or suppresses a non-minimal coupling between these components (see \cite{carroll} for a discussion), a small interaction cannot be ruled out.  

The usual critique to coupled quintessence scenarios is that, in the absence of a natural guidance from fundamental physics, one needs to specify a possible interacting or coupling term between the two dark components in order to establish a model and investigate their observational and theoretical implications.  In this concern, a still phenomenological but very interesting step toward a more realistic interacting or coupling law was recently discussed in Ref.~\cite{wm} (see also~\cite{alclim05}) in the context of models with vacuum decay $(\omega =-1)$. Instead of the traditional approach, Ref.~\cite{wm} deduced the new interaction law from a simple argument about the effect of the dark energy on the CDM expansion rate. Such a coupling is similar to the one obtained in Ref.~\cite{shapiro} from arguments based on renormalization group and seems to be very general, having many of the previous attempts as a particular case.

An important aspect worth emphasizing is that in the above analyses the interacting parameter $\epsilon$ has been considered constant over the cosmic evolution whereas in a more realistic case it must be a time-dependent quantity. In Ref.~\cite{ernandes2} the above arguments were extend for the case in which the interacting parameter $\epsilon$ is a function of the scale factor $a$. The analysis of Ref.~\cite{ernandes2}, however, was restricted to the case in which $\omega = -1$, which is mathematically equivalent to dynamical $\Lambda$ scenarios. 

In this paper, we extend the arguments of Ref.~\cite{ernandes2} to a dark energy/dark matter interaction, where the dark energy component is described by an equation of state $p_{x}=\omega \rho_{x}$ with $\omega < 0$. We explore the dynamical behavior of this class of models and find viable cosmological solutions for two parameterizations of $\epsilon(a)$. In particular, the solutions with transient acceleration, as consequence of the interaction in the dark sector, are investigated in more detail. In order to check the observational viability of this general class of coupled quintessence scenarios, we also carry out a statistical analysis with recent observations of type Ia supernovae (SNe Ia) along with recent estimates of the CMB/BAO ratio at two different redshifts $z = 0.20$ and $z = 0.35$.

\section{Interaction in the dark sector}

\begin{figure*}
\centerline{\psfig{figure=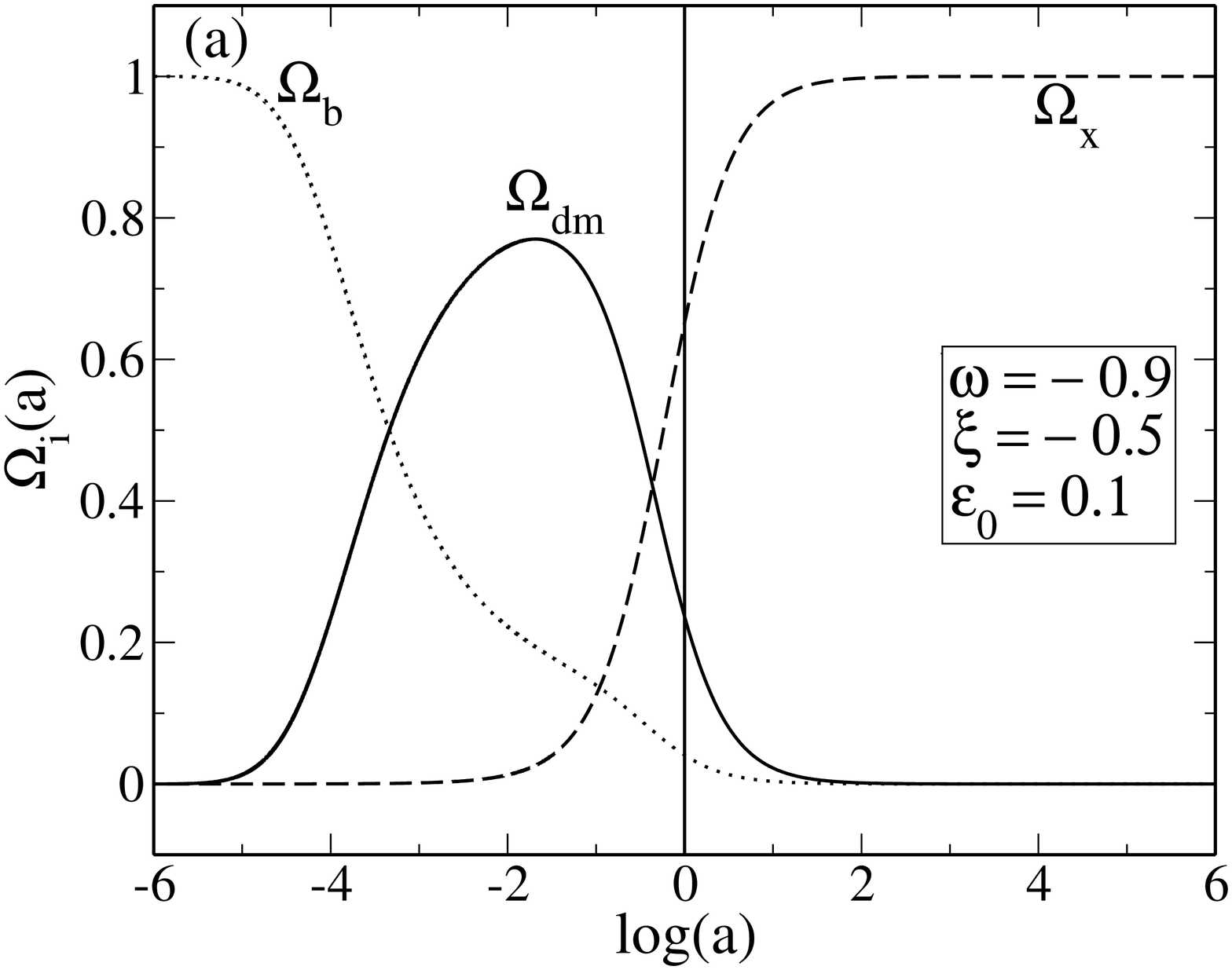,width=2.0truein,height=1.9truein,angle=-90} 
\psfig{figure=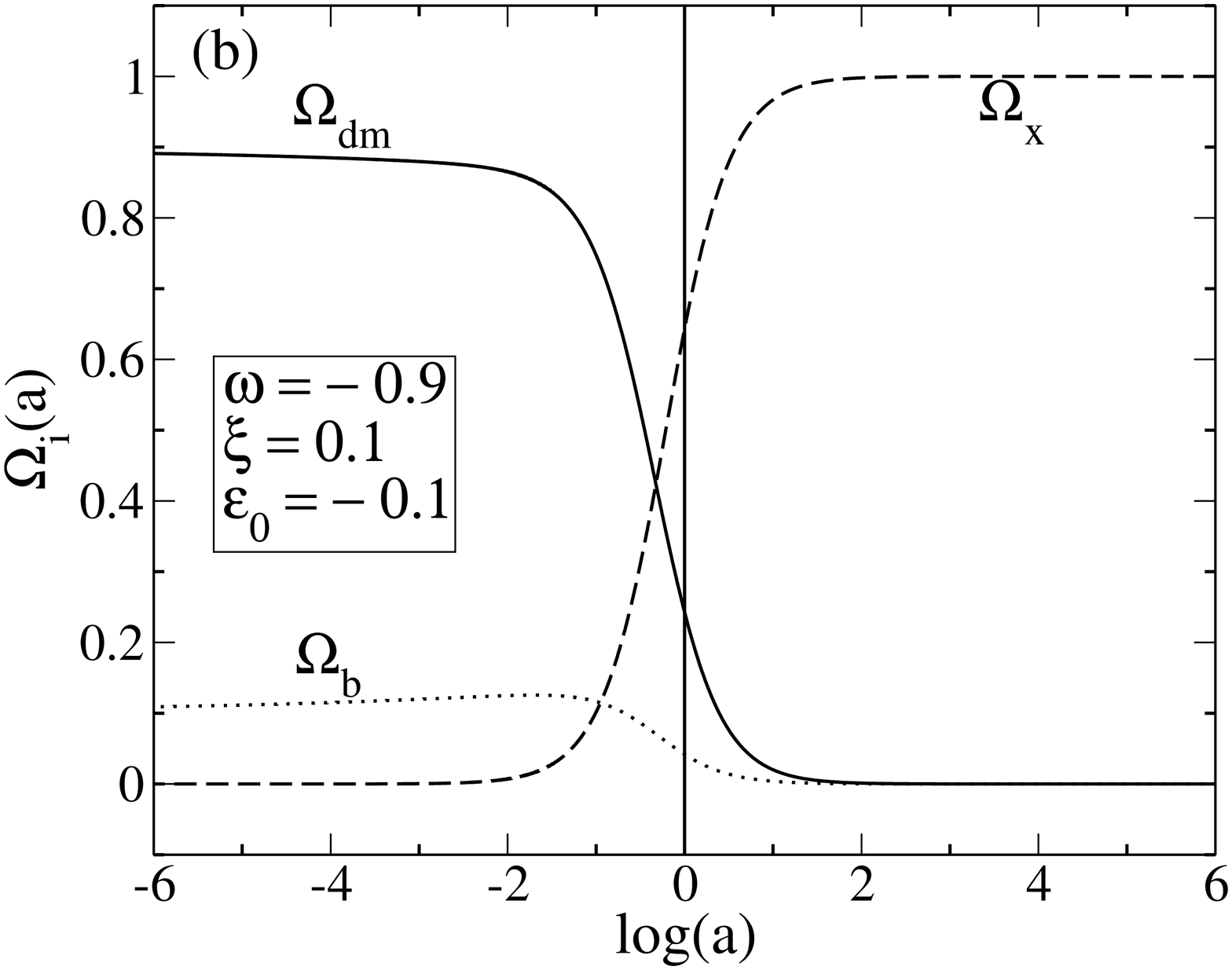,width=2.0truein,height=1.9truein,angle=-90}
\psfig{figure=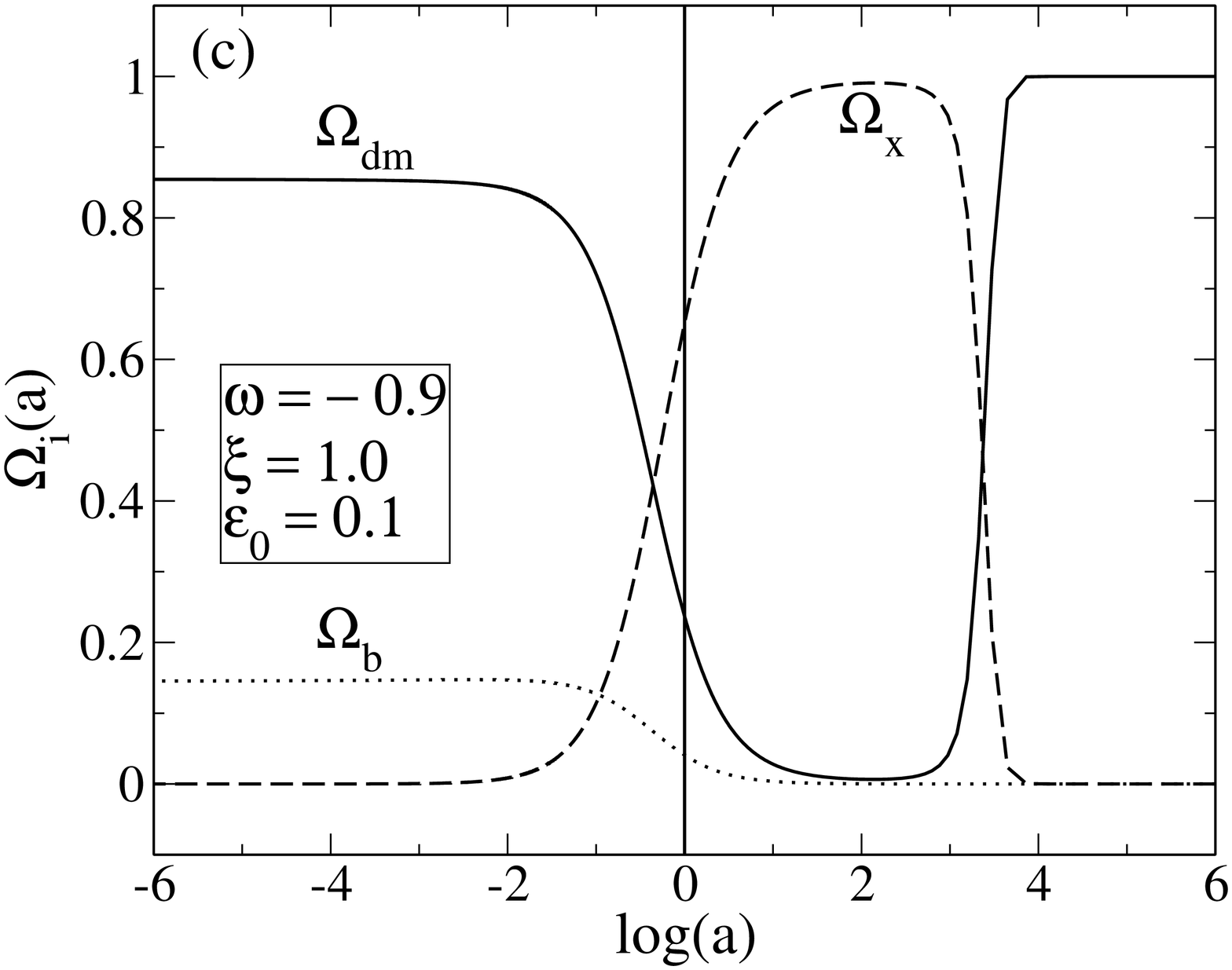,width=2.0truein,height=1.9truein,angle=-90}} 
\label{fig:qzw}
\end{figure*}
\begin{figure*}
\centerline{\psfig{figure=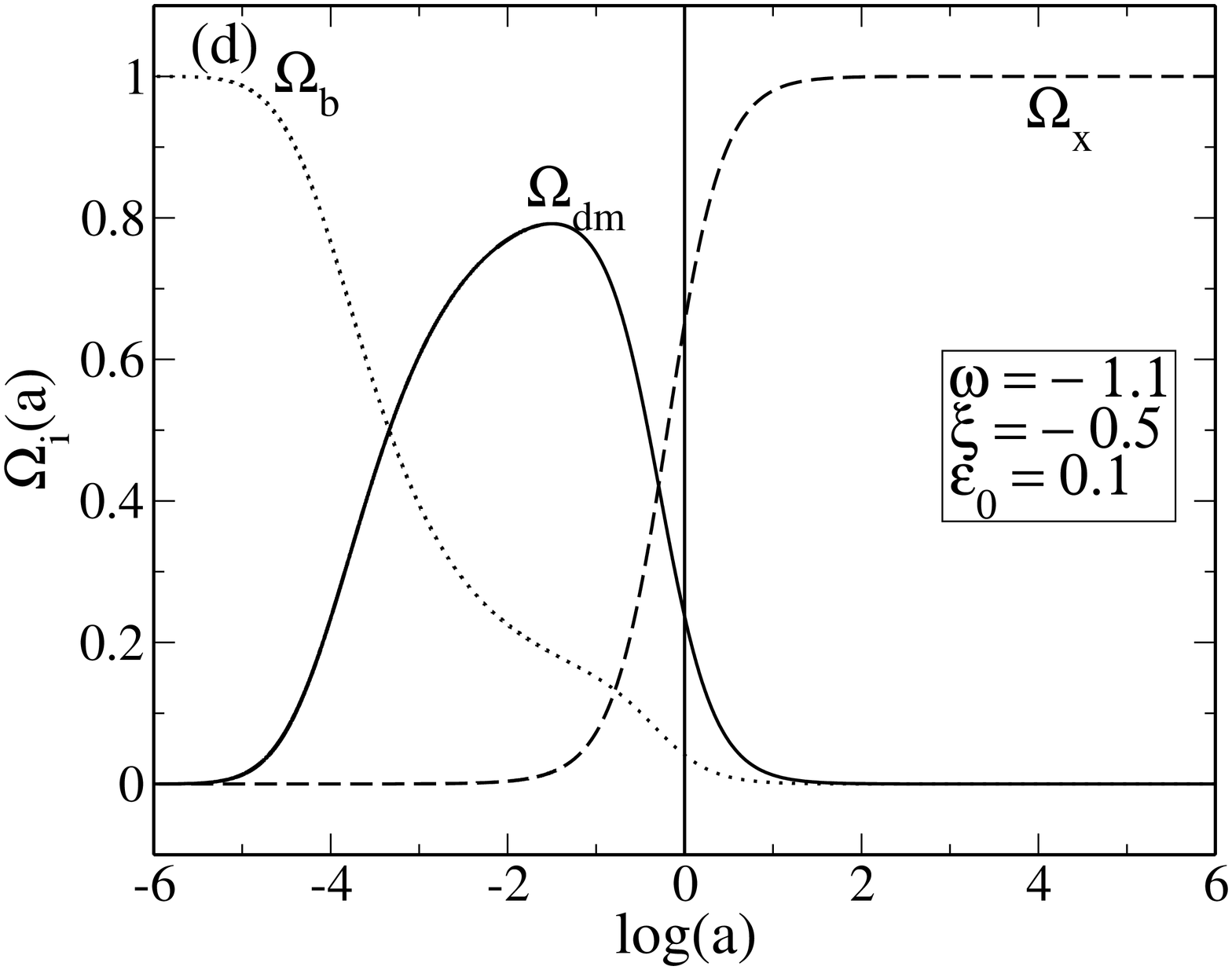,width=2.0truein,height=1.9truein,angle=-90} 
\psfig{figure=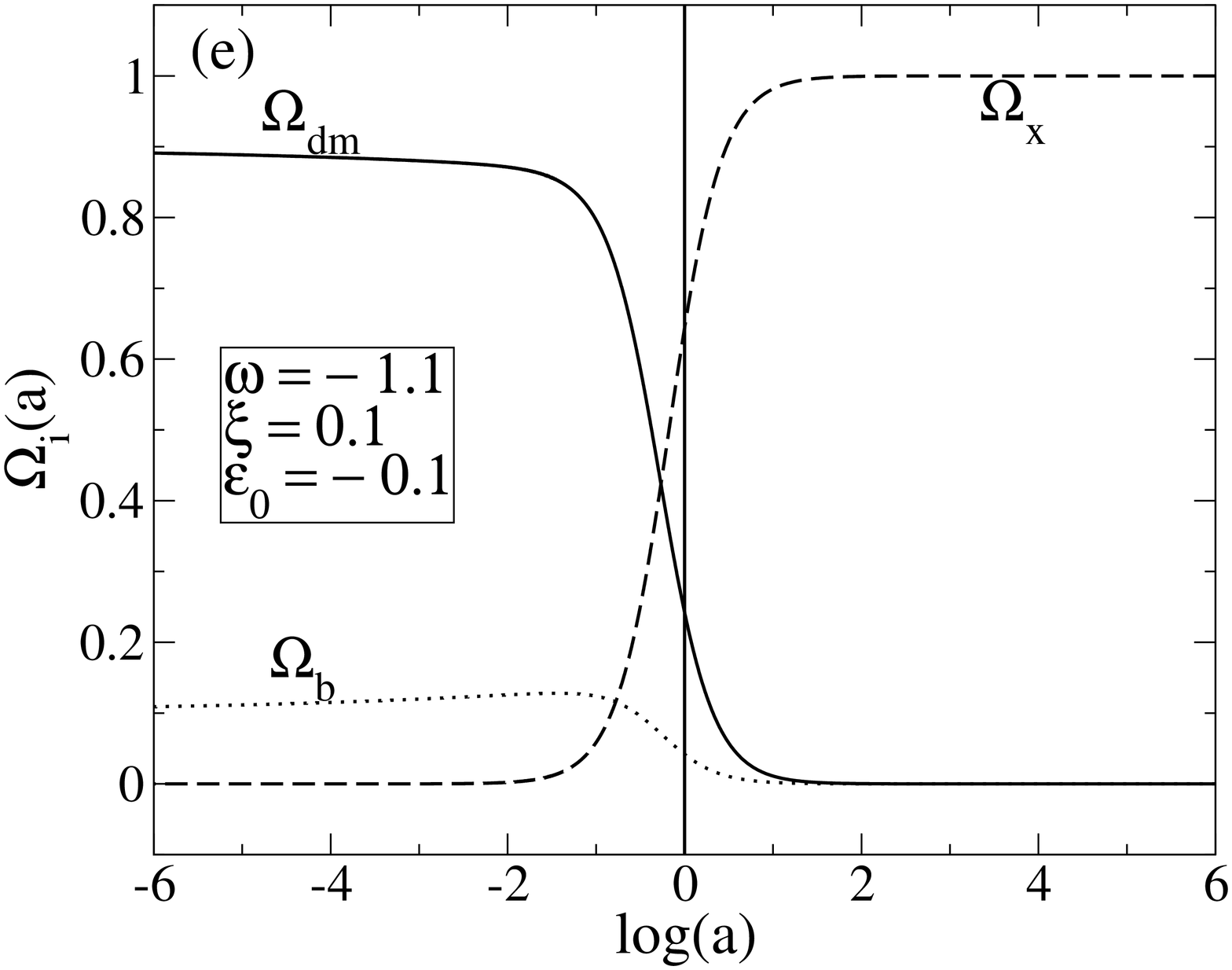,width=2.0truein,height=1.9truein,angle=-90}
\psfig{figure=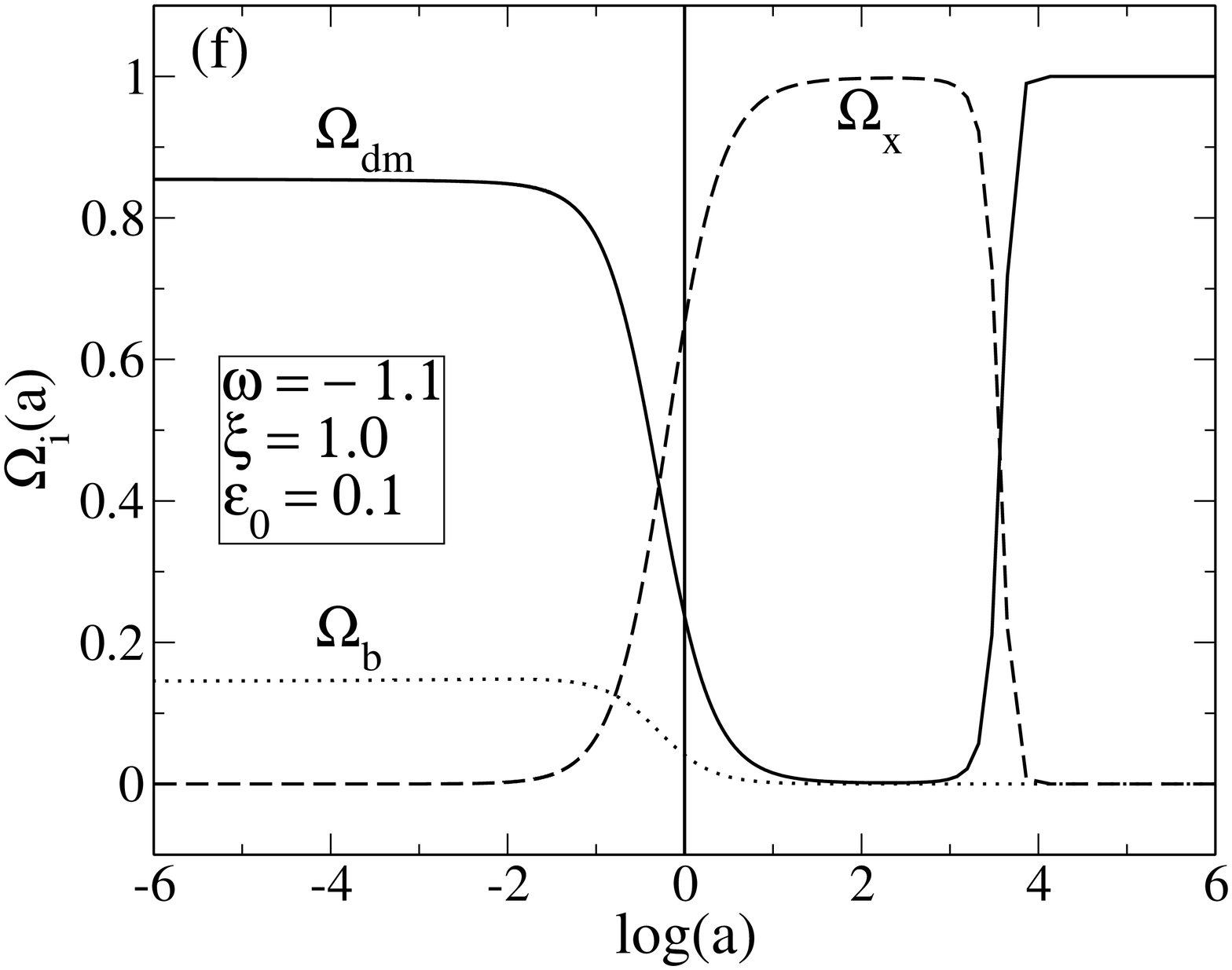,width=2.0truein,height=1.9truein,angle=-90}} 
\caption{Evolution of the density parameters $\Omega_i$ ($i = b, dm, x$) as a function of $\log(a)$ for some selected combinations of $\epsilon_{0}$ and $\xi$, [Eq. (\ref{para1})] and two characteristic values of the equation-of-state parameter, $w = -0.9$ and $w = -1.1$, corresponding to quintessence and phantom behaviors, respectively. The case $\omega = -1.0$ was discussed in Ref.~\cite{ernandes2}.}
\label{fig:qzw}
\end{figure*}

According to current observations, the main contributions to the total energy-momentum tensor of the cosmic fluid are non-relativistic matter (baryonic plus dark) and a negative-pressure dark energy component. By assuming that a possible interaction occurs in the dark sector, the energy conservation equation for the two interacting components can be written as
\begin{equation}\label{coup}
\dot{\rho}_{dm} + 3 \frac{\dot{a}}{a}\rho_{dm} = -\dot{\rho}_x -
3\frac{\dot{a}}{a}(\rho_x + p_x)\;,
\end{equation}
where $\rho_{dm}$ and $\rho_x$, are the energy densities of the dark matter and dark energy, respectively, whereas $p_{x}$ is the dark energy pressure. As the dark components are exchanging energy, dark matter density will dilute in a rate whose deviation from standard case, $\rho_{dm} \propto a^{-3}$, may be characterized by the function $\epsilon(a)$, i.e.,
\begin{equation}
\label{dm}
\rho_{dm}=\rho_{dm,o} a^{-3 + \epsilon(a)},
\end{equation} 
where the subscript 0 denotes current values and we have set $a_{0}=1$. Note that, contrarily to most analyses available in the literature, we consider the interaction parameter as a function of the cosmological scale factor, $\epsilon = \epsilon(a)$. In what follows we also consider that the dark energy is described by an equation of state $p_x=\omega \rho_x$, with $\omega = \rm{constant} < 0$. 

By substituting the above evolution law into Eq. (\ref{coup}), we find 
\begin{equation}\label{de}
\rho_{x} =  \left[\rho_{x,0} + \rho_{dm,0} \int_{a}^{1}\frac{[\epsilon(\tilde{a}) + \tilde{a} \epsilon^{'} ln \tilde{a}]}{\tilde{a}^{1 -3\omega - \epsilon(\tilde{a})}} d\tilde{a} \right]a^{-3(1+\omega)},
\end{equation}
where a prime denotes derivative with respect to scale factor $a$. For $\omega=-1$ the above equation reduces to the vacuum decaying scenario recently discussed in Ref. \cite{ernandes2}, whereas for $\omega \neq - 1$ and $\epsilon = \rm{const.}$, the above expressions reduce to the scenario recently discussed in Refs.~\cite{jesus, ernandes1}.

\begin{figure*}
\centerline{\psfig{figure=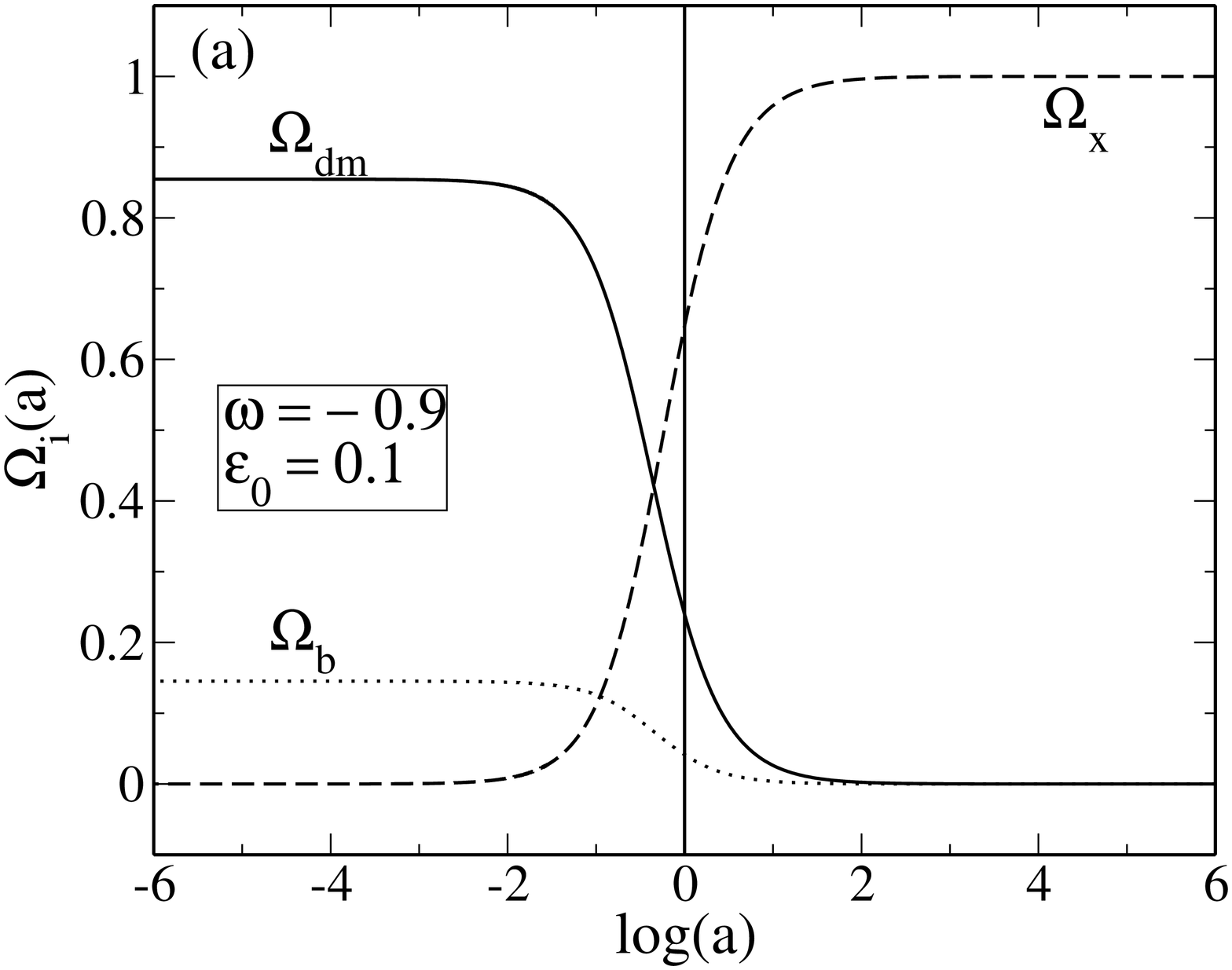,width=1.7truein,height=1.8truein,angle=-90} 
\psfig{figure=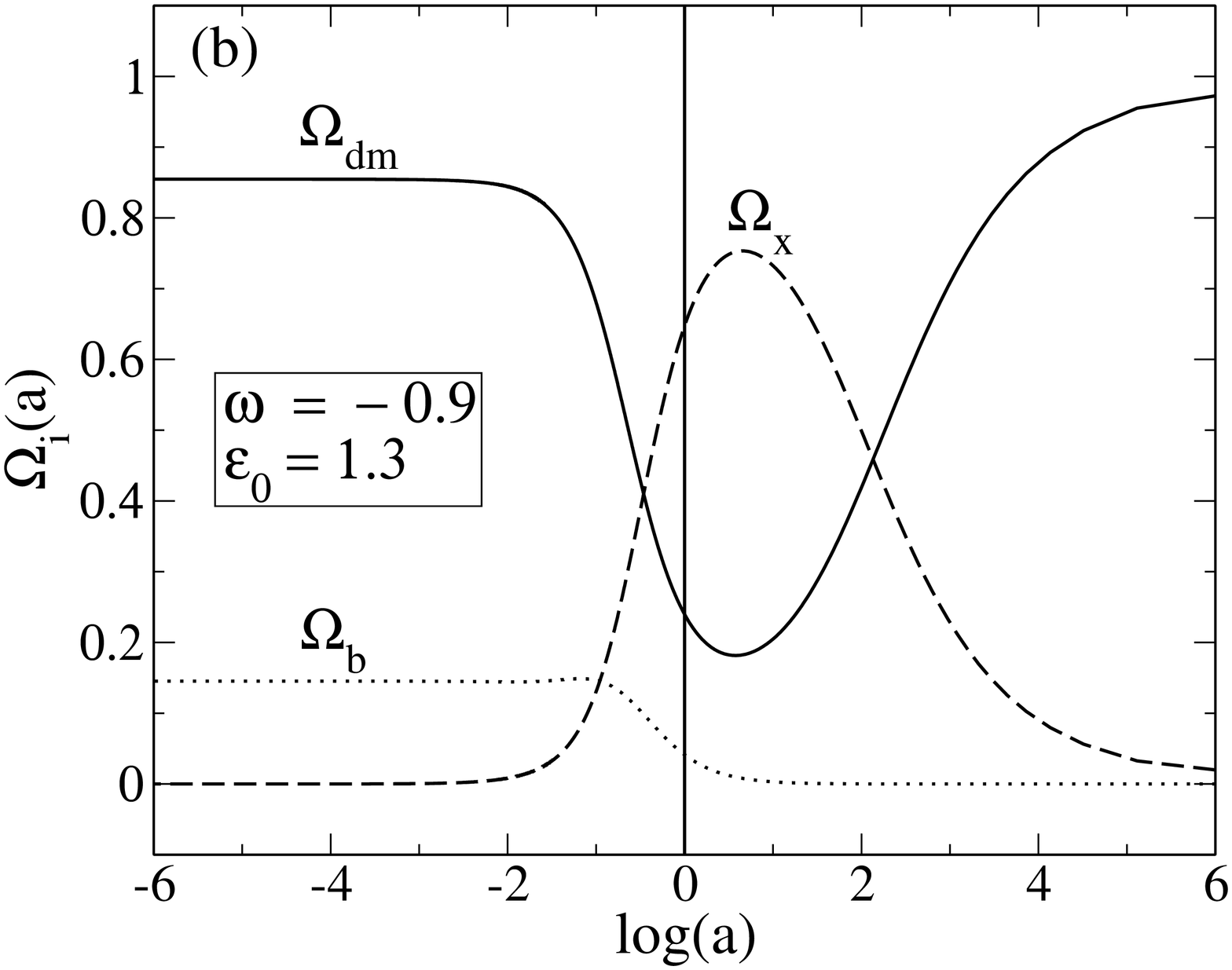,width=1.7truein,height=1.8truein,angle=-90}
\psfig{figure=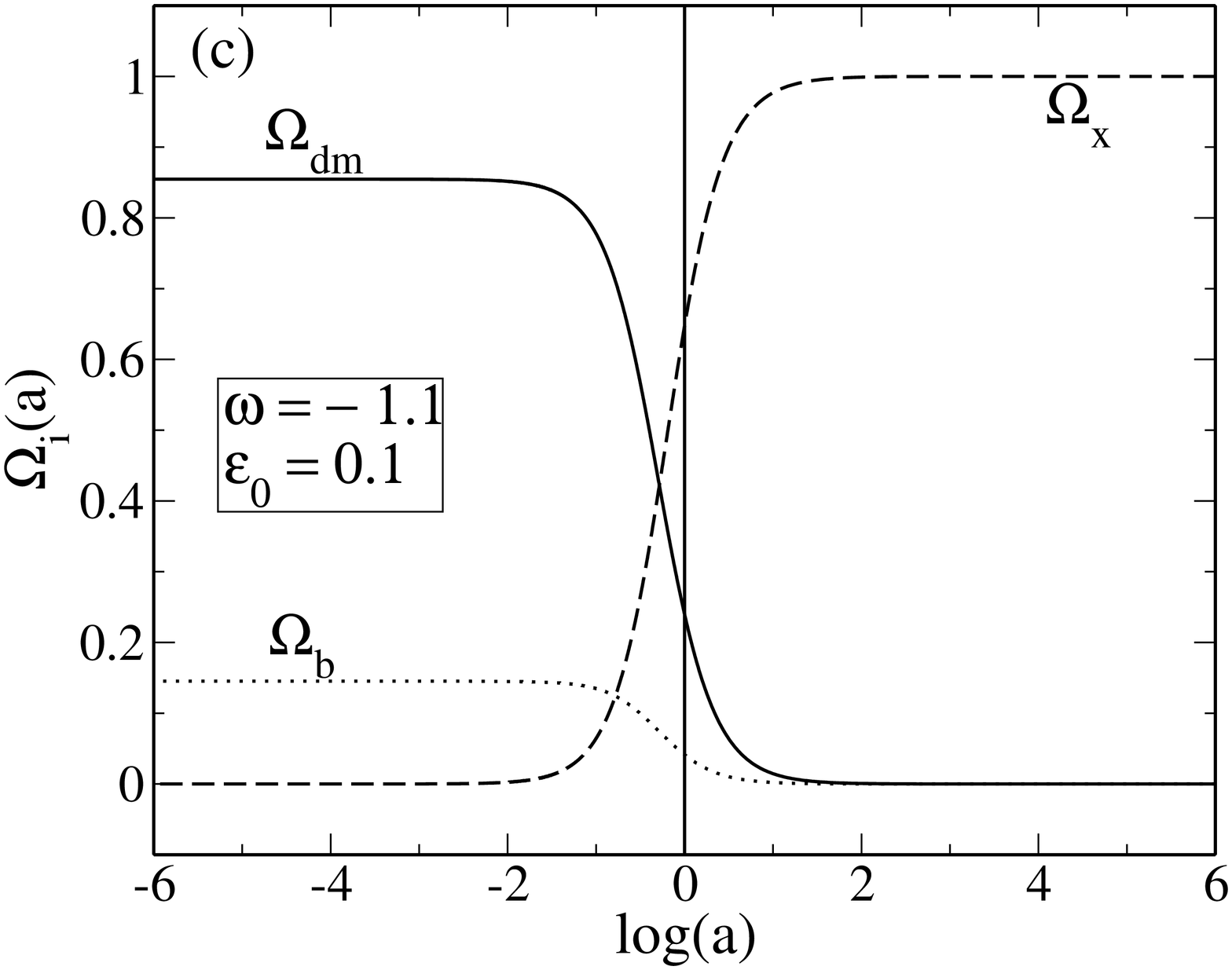,width=1.7truein,height=1.8truein,angle=-90}
\psfig{figure=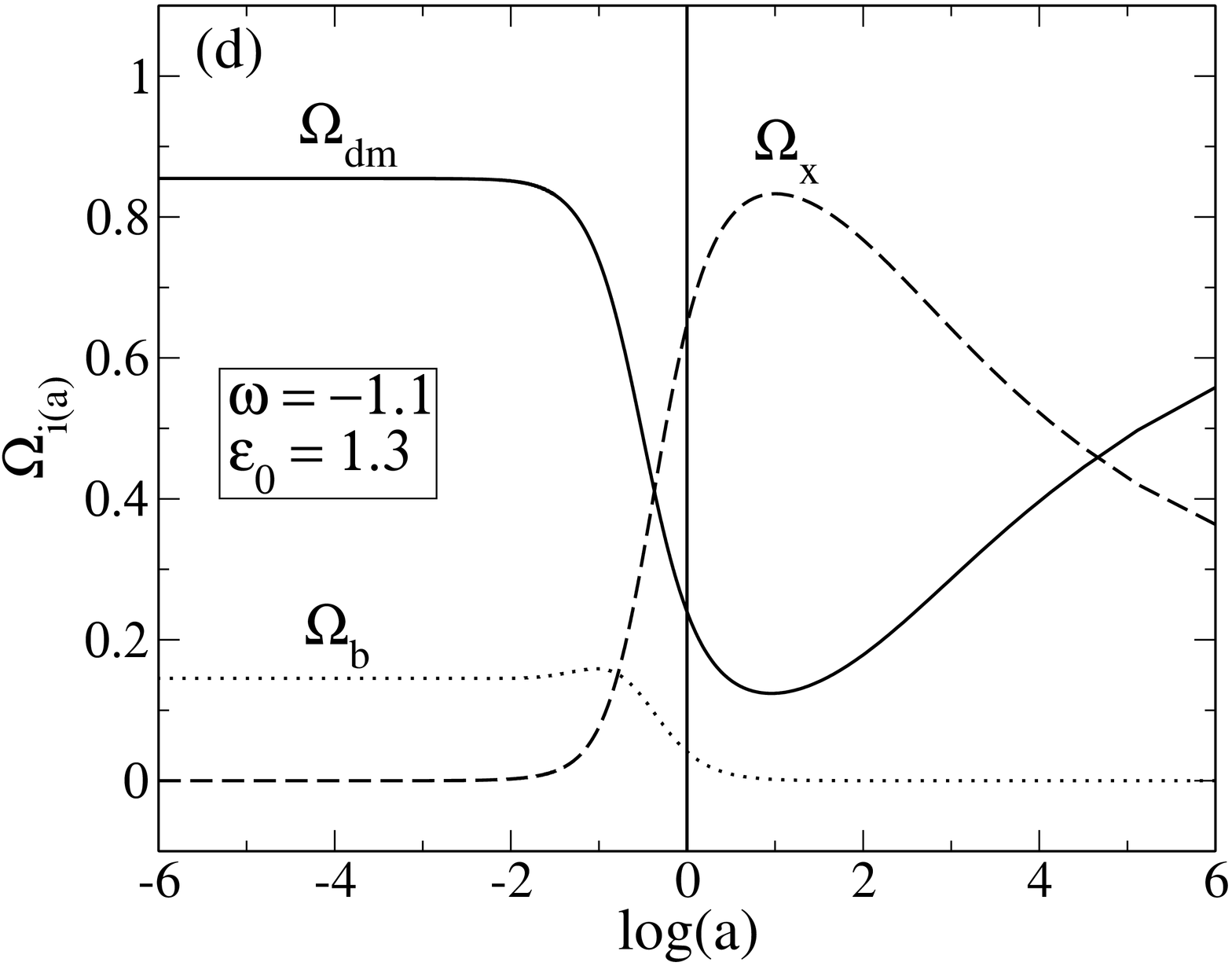,width=1.7truein,height=1.8truein,angle=-90}} 
\caption{The same as in Fig. 1 for some selected values of $\epsilon_{0}$, [Eq. (\ref{para2})]. Note that for large positive values of $\epsilon_{0} > 1.2$ (Panel 2b and 2d), the interaction between dark matter and dark energy will drive the Universe to a new matter-dominated era in the future, when $a \rightarrow \infty$.}
\label{fig:qzw}
\end{figure*}

We assume from now on vanishing spatial curvature (and neglect the radiation contribution), so that the Friedmann equation for this interacting dark matter-dark energy scenario can be written as
\begin{equation}\label{friedmann}
{\cal{H}} = \left[ \Omega_{b,0} a^{-3} + \Omega_{dm,0}a^{-3 + \epsilon(a)} + \Omega_{x,0}f(a) \right]^{1/2}\;,
\end{equation}
where ${\cal{H}}={{H}}/{H_o}$, and $\Omega_{b,0}$, $\Omega_{dm,0}$ and $\Omega_{x,0}$ stand for, respectively, the current baryon, dark matter and dark energy density parameters. In the above equation, the dimensionless function $f(a)$ takes the following form:
\begin{equation}\label{fdea}
f(a) = \left[1+ \frac{\Omega_{dm,0}}{\Omega_{x,0}}\int_{a}^{1}\frac{[\epsilon(\tilde{a}) + \tilde{a} \epsilon^{'} \ln \tilde{a}]}{\tilde{a}^{1 -3\omega - \epsilon(\tilde{a})}} d\tilde{a} \right]a^{-3(1+\omega)}.
\end{equation}

\subsection{$\epsilon(a)$ parameterization}

In order to proceed further and study some cosmological consequences of the class of coupled quintessence scenarios discussed above, we must assume an appropriated relation for $\epsilon(a)$. In our analysis, we consider two different parameterizations for the interacting parameter: 
\begin{equation}\label{para1}
\epsilon(a) = \epsilon_0a^\xi = \epsilon_0(1+z)^{-\xi}, \quad \quad \quad \quad \quad \rm{(P1)}\\\;
\end{equation}
and
\begin{equation}
\label{para2}
\epsilon (a) = \epsilon_{0} \exp{(1 -a^{-1})} = \epsilon_0\exp{(-z)}, \quad \rm{(P2)}\\\;
\end{equation}
where $\epsilon_0$ and $\xi$ may, in principle, take negative and positive values. P1 is certainly a very simple choice among some physically possible functional forms. Note, however, that for negative values of $\xi$, P1 blows up in the past, when $a \rightarrow 0$. Differently, P2 is a one-parameter, well-behaved function during the entire evolution of the Universe. Note also that P2 implies a weaker dark matter/dark energy interaction in the past, as $z$ increases.

\section{Dynamical behavior}

The time evolution of the density parameters $\Omega_b(a)$, $\Omega_{dm}(a)$ and $\Omega_{x}(a)$ can be derived by combining Eqs. (\ref{dm})-(\ref{friedmann}). They read:
\begin{subequations}
\begin{equation} \label{8a}
\Omega_{b}(a) = \frac{a^{-3}}{a^{-3} + {\rm{A}}a^{-3 +\epsilon(a)} + Bf(a)}\;,
\end{equation}
\begin{equation} \label{8b}
\Omega_{dm}(a) = \frac{a^{-3 +\epsilon(a)}}{A^{-1}a^{-3} + a^{-3 +\epsilon(a)} + Cf(a)}\;,
\end{equation}
\begin{equation} \label{8c}
\Omega_{x}(a) = \frac{f(a)} {B^{-1}a^{-3} + C^{-1}a^{-3 + \epsilon(a)} + f(a)}\;,
\end{equation}
\end{subequations}
where  $A = \Omega_{dm,0}/{\Omega_{b,0}}$, ${\rm{B}} = {\Omega_{x,0}}/{\Omega_{b,0}}$ and  ${\rm{C}} = {\Omega_{x,0}}/{\Omega_{dm,0}}$. 

Figure 1 shows the evolution of the density parameters as function of $\log(a)$ for P1 [Eq. (\ref{para1})]. For simplicity, we consider two characteristic values of the equation-of-state parameter, $w = -0.9$ and $w = -1.1$, corresponding to quintessence and phantom behaviors, respectively. In agreement with current WMAP results~\cite{cmbnew}, we assume $\Omega_{b,0} = 0.0416$ and $\Omega_{dm,0} = 0.24$. Note that, although currently accelerated (and, therefore, possibly in agreement with SNe Ia data), models with $\epsilon_{0} > 0$ and negative values of $\xi$  (Figs. 1a and 1d) fail to reproduce the past dark matter-dominated epoch, whose existence is fundamental for the structure formation process to take place. In both cases, the dark energy and dark matter densities vanish at high-$z$ and the Universe is fully dominated by the baryons (for a CMB analysis in a baryon-dominated universe, see~\cite{silk}).

\begin{figure*}
\centerline{\psfig{figure=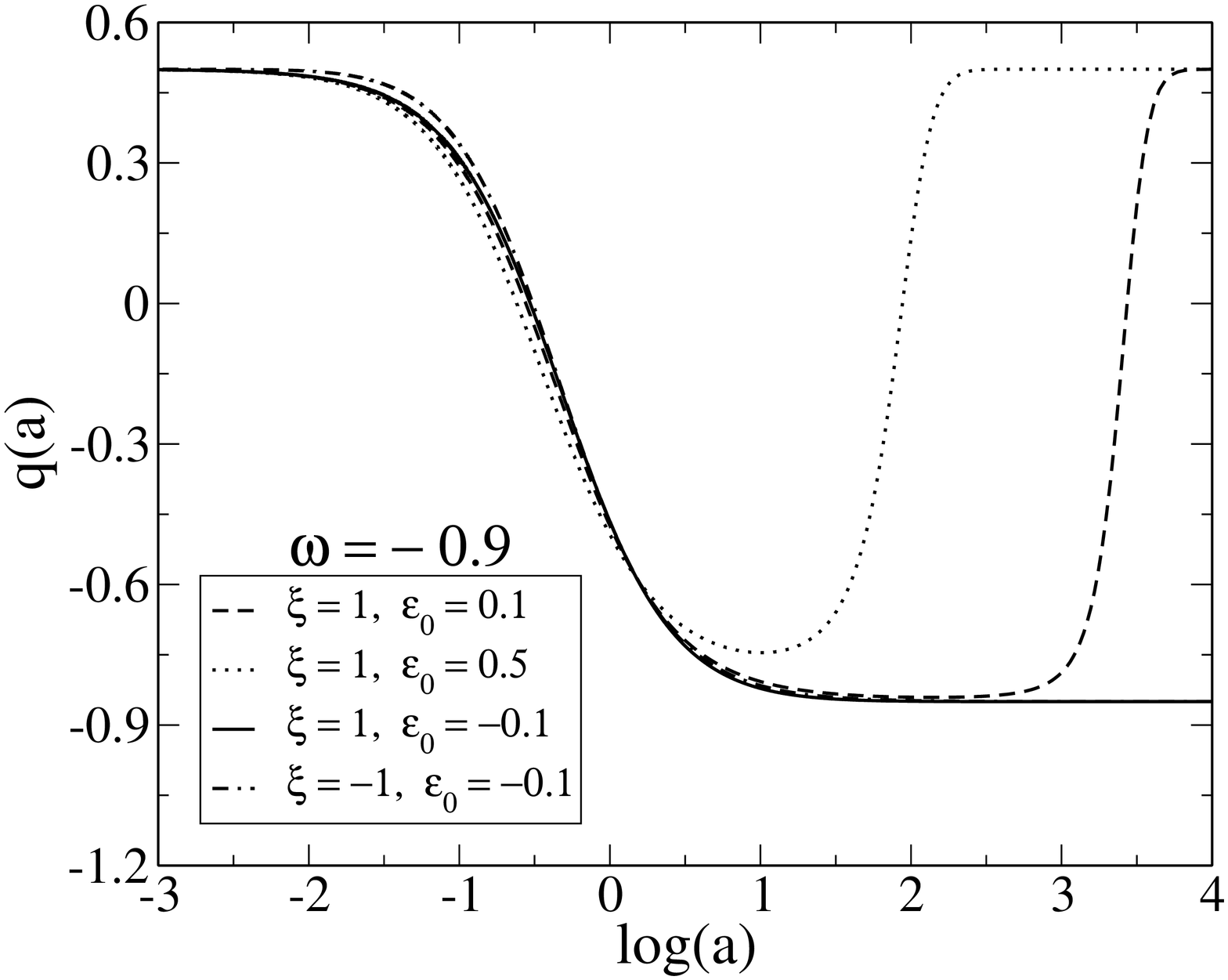,width=2.1truein,height=1.9truein,angle=-90}
\psfig{figure=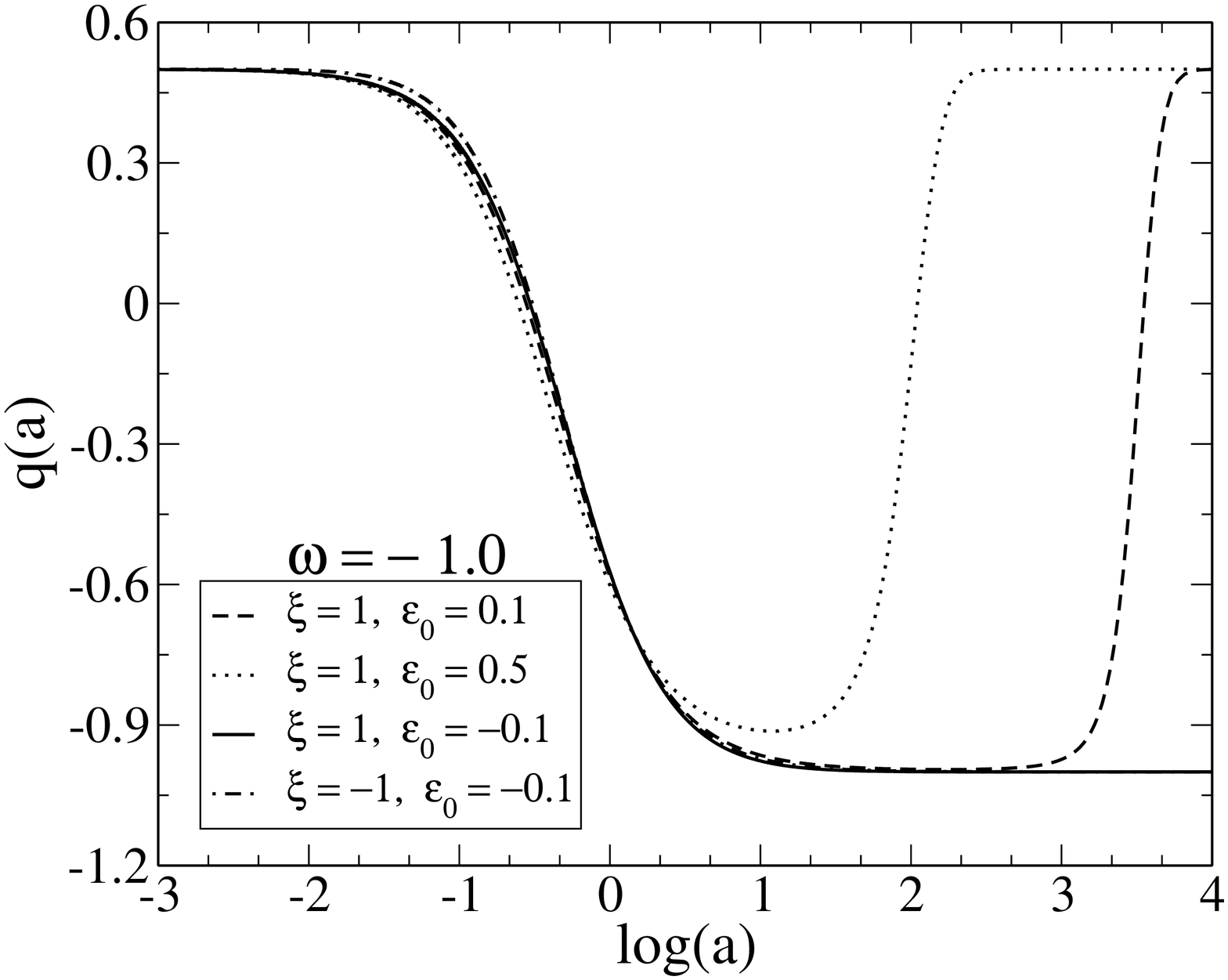,width=2.1truein,height=1.9truein,angle=-90} 
\psfig{figure=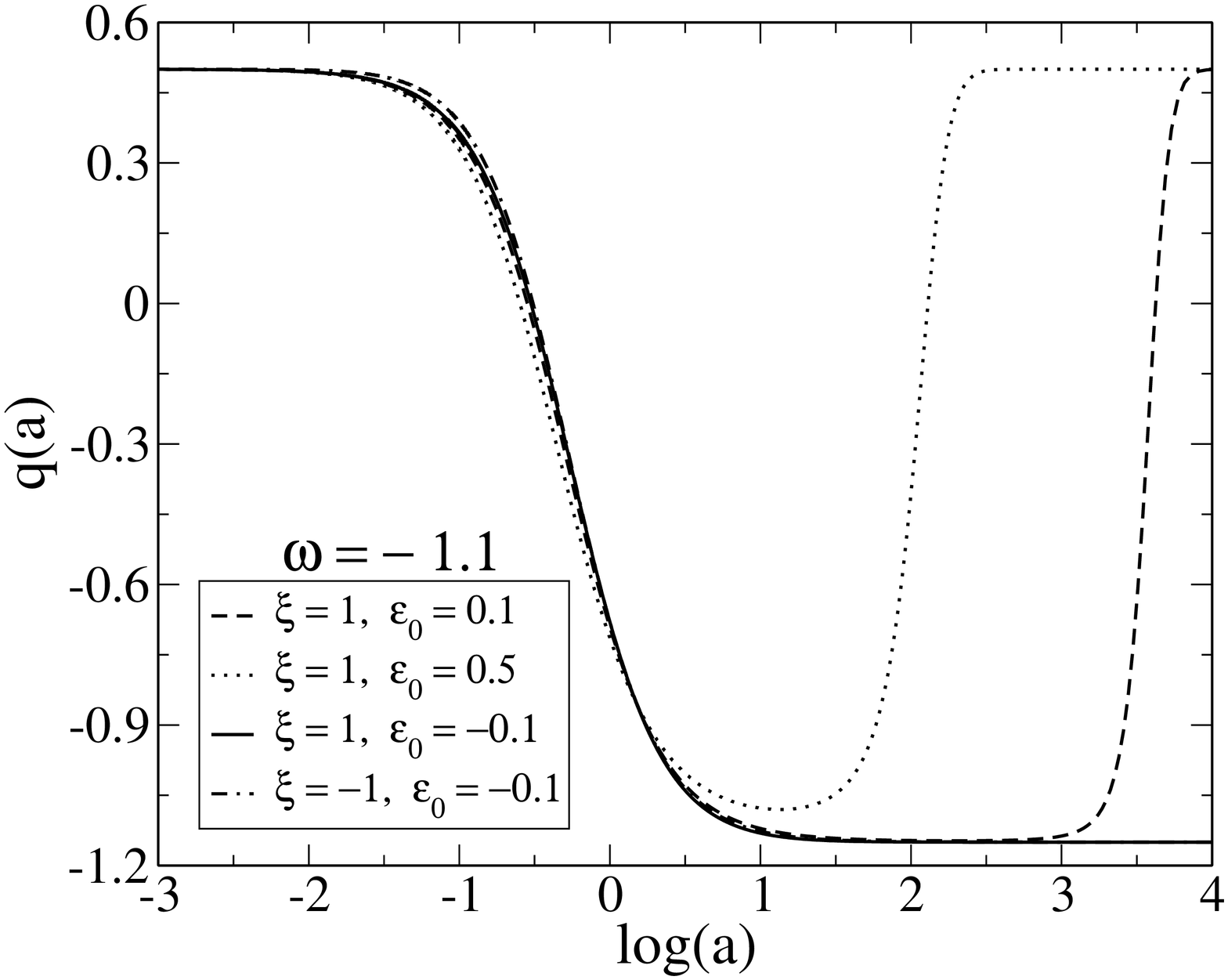,width=2.1truein,height=1.9truein,angle=-90}}
\label{fig:qzw}
\end{figure*}
\begin{figure*}
\centerline{\psfig{figure=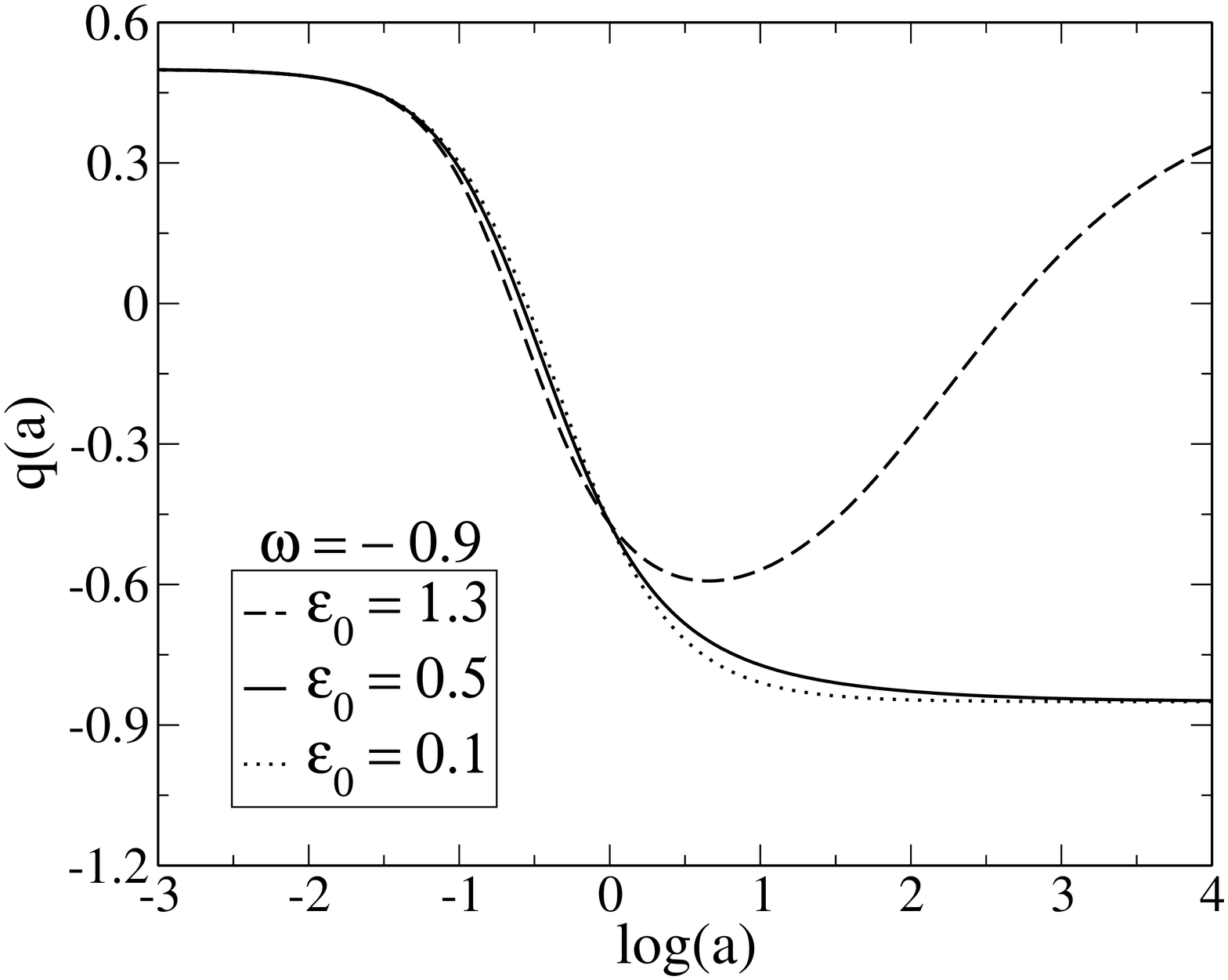,width=2.1truein,height=1.9truein,angle=-90}
\psfig{figure=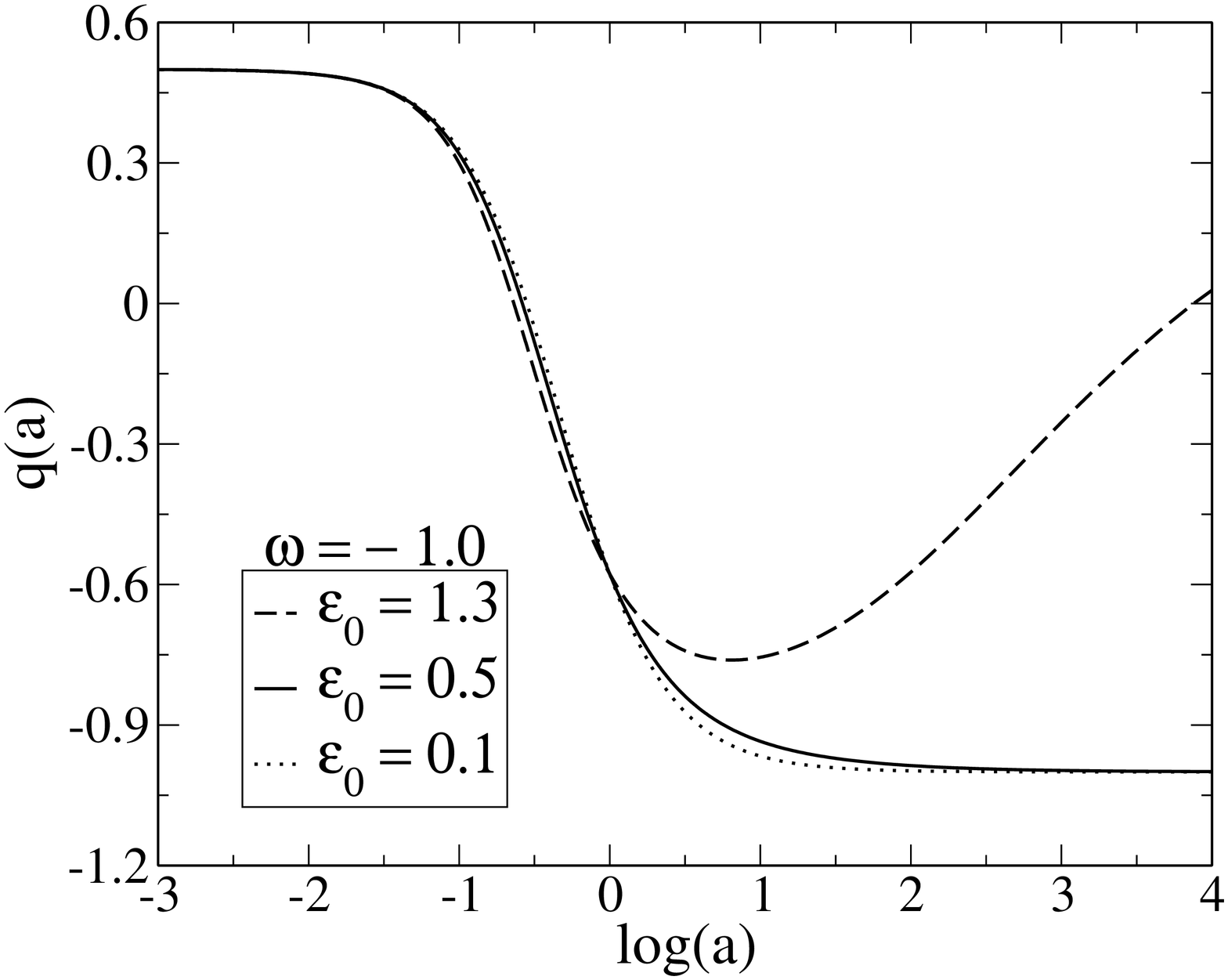,width=2.1truein,height=1.9truein,angle=-90} 
\psfig{figure=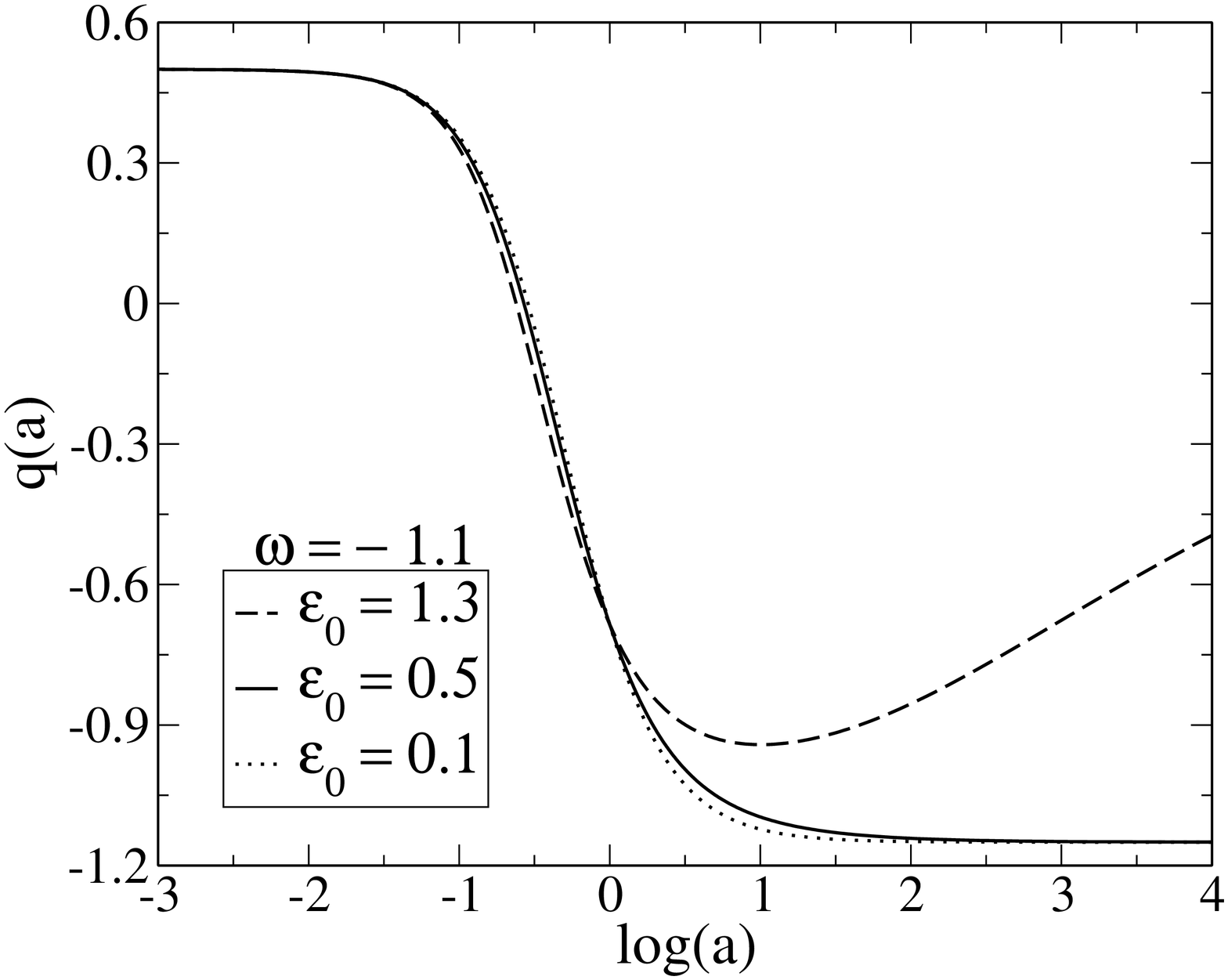,width=2.1truein,height=1.9truein,angle=-90}} 
\caption{Deceleration parameter as a function of $\log(a)$ for some selected values of $\epsilon_{0}$ and $\xi$ for P1 and P2, respectively. Note that for $\epsilon_{0} > 0$ and large positive values of $\xi$ (P1) and also when $\epsilon_{0} > 1.2$ (P2) the Universe will experience a new matter-dominated era in the future, when $a \rightarrow \infty$.}
\label{fig:qzw}
\end{figure*}

Regardless of the sign of $\epsilon_0$ and the values of $w$, well-behaved scenarios are obtained when $\xi$ takes positive values (Figs. 1b and 1e). In these cases  a mix of baryons ($\lesssim 20\%$) and dark matter ($\gtrsim 80\%$) dominates the past evolution of the Universe whereas the dark energy is always the dominant component from a value of $a \lesssim 1$ on. An interesting and completely different future cosmic evolution is obtained when $\epsilon_0 > 0$ and the parameter $\xi$ takes large positive values ($\gtrsim 0.8$). This is shown in (Figs. 1c and 1f) for $\xi = 1.0$ and $\epsilon_0 = 0.1$. Note that, besides having a well-behaved past evolution and being currently accelerating, the cosmic acceleration will eventually stop at some value of $a>>1$ (when the dark energy becomes sub-dominant) and the Universe will experience a new matter-dominated era in the future, when $a \rightarrow \infty$. This kind of dynamic behavior is not  found in most coupled quintessence models discussed in the literature, being essentially a feature of the so-called thawing~\cite{thaw} and hybrid~\cite{cqg} potentials, which in turn seems to be in good agreement with some requirements of String or M theories, as discussed in Ref.~\cite{fischler} (see also \cite{ed})~\footnote{The argument presented in Ref.~\cite{fischler} is that an eternally accelerating universe, a rather generic feature of many quintessence scenarios (including the standard $\Lambda$CDM model), seems not to be in agreement with String/M-theory predictions, since it is endowed with a cosmological event horizon which prevents the construction of a conventional S-matrix describing particle interactions.}.

In Fig. 2 it is shown the same as Fig. 1 for P2 [Eq. (\ref{para2})]. Note that, independently of the values of $\omega$, all values of $\epsilon_{0} > 0$ give rise to well-behaved scenarios in which the Universe had a last evolution dominated by a mix of baryons  and dark matter  and it is currently accelerating (dominated by dark energy). Note also that for large positive values $\epsilon_{0}$, e.g., $\epsilon_{0} \gtrsim 1.2$ (Panels 2b and 2d) the Universe will evolve to an eternal deceleration phase instead of the usual de Sitter phase. 

The deceleration parameter, defined as $q=-a\ddot{a}/\dot{a}^2$, is given by
\begin{equation}\label{acela}
q= \frac{3}{2}{\frac{\Omega_{b,0}a^{-3} + \Omega_{dm,0}a^{-3+\epsilon(a)} + (1+\omega) \Omega_{x,0}f(a)} {\Omega_{b,0}a^{-3} + \Omega_{dm,0}a^{-3+\epsilon(a)} + \Omega_{x,0}f(a)}} -1,
\end{equation}
and shown in Fig. 3 as a function of $\log(a)$ for P1 and P2, respectively. In agreement with our previous discussion, we clearly see a transient acceleration phenomenon for some selected values of $\epsilon_0$ and $w$.

\begin{figure*}
\centerline{\psfig{figure=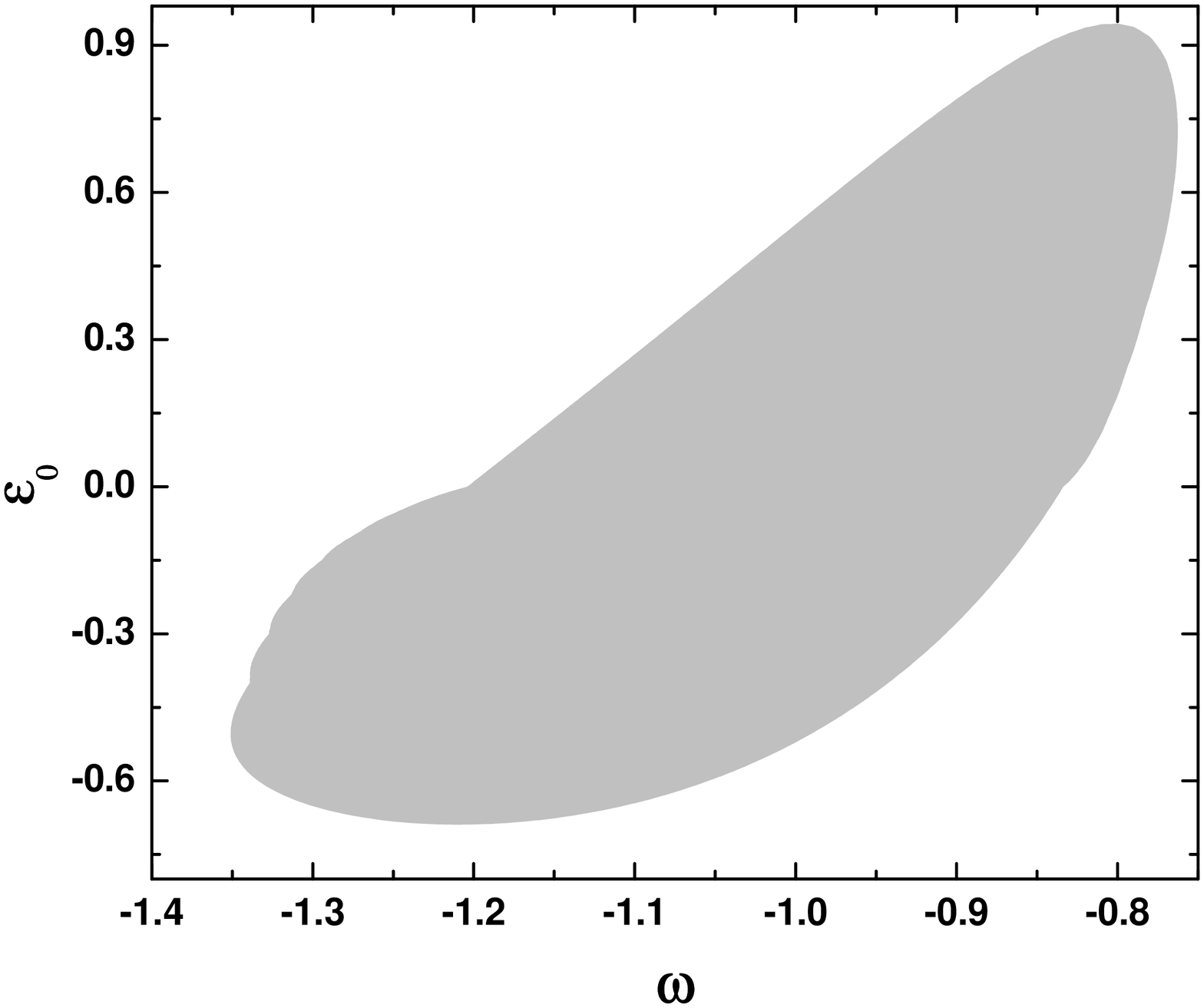,width=3.2truein,height=2.0truein,angle=-90} 
\psfig{figure=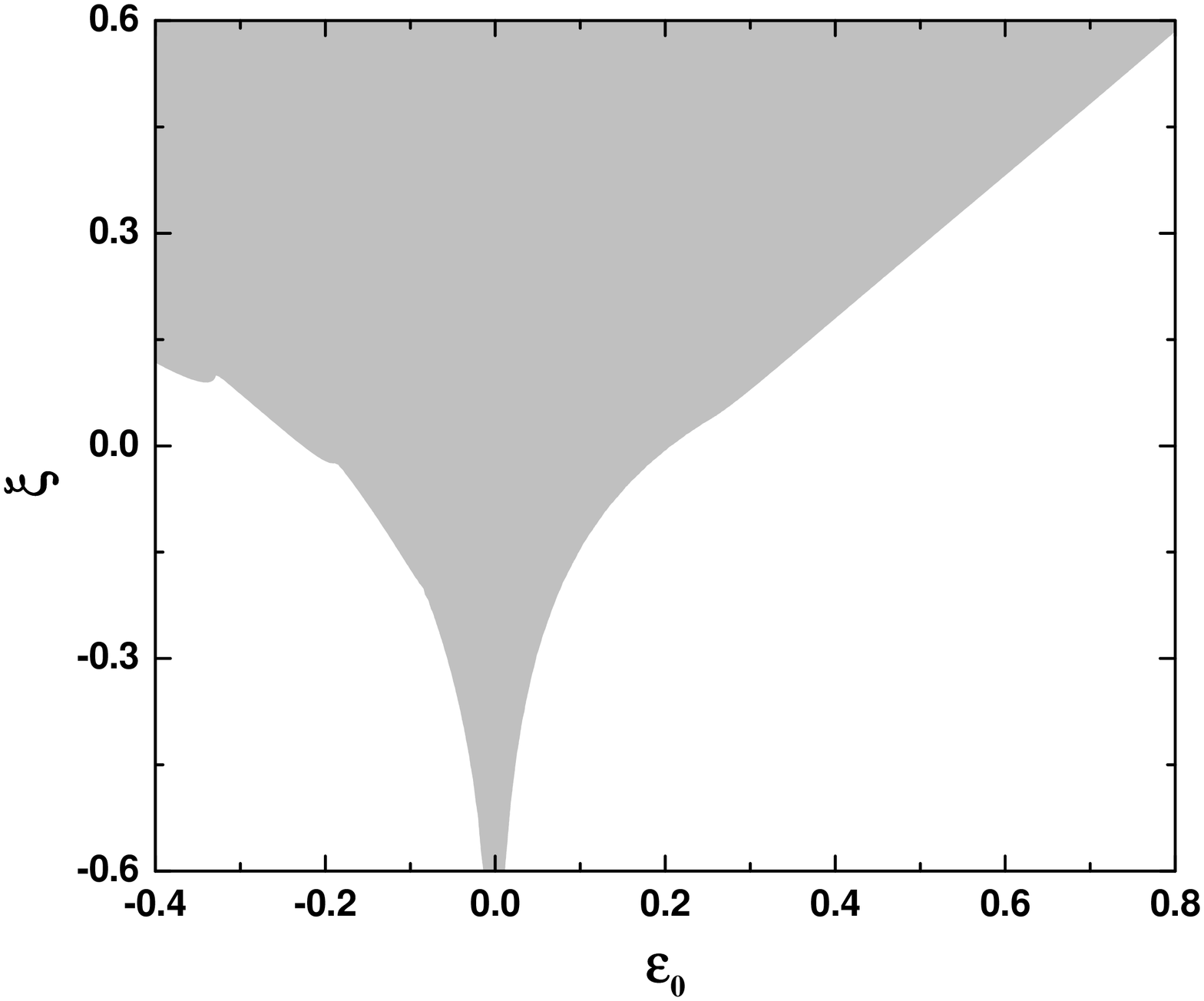,width=3.2truein,height=2.0truein,angle=-90}}
\caption{The results of our statistical analyses. Contours of $\chi^2$ in the plane $\omega - \epsilon_0$ (left panel) and $\epsilon_0 - \xi$, with $\omega = -1$, (right panel) for P1. These contours are drawn for $\Delta \chi^2 = 2.30$ and $6.17$. The best fit values are $\epsilon_{0}= -0.11$ and $\omega = -1.04$ (left panel), while in the case $\Lambda(t)$CDM (right panel) we have found $\epsilon_{0}= 0.49$ and $\xi = 0.41$.}
\label{fig:qzw}
\end{figure*}

\begin{figure*}
\centerline{\psfig{figure=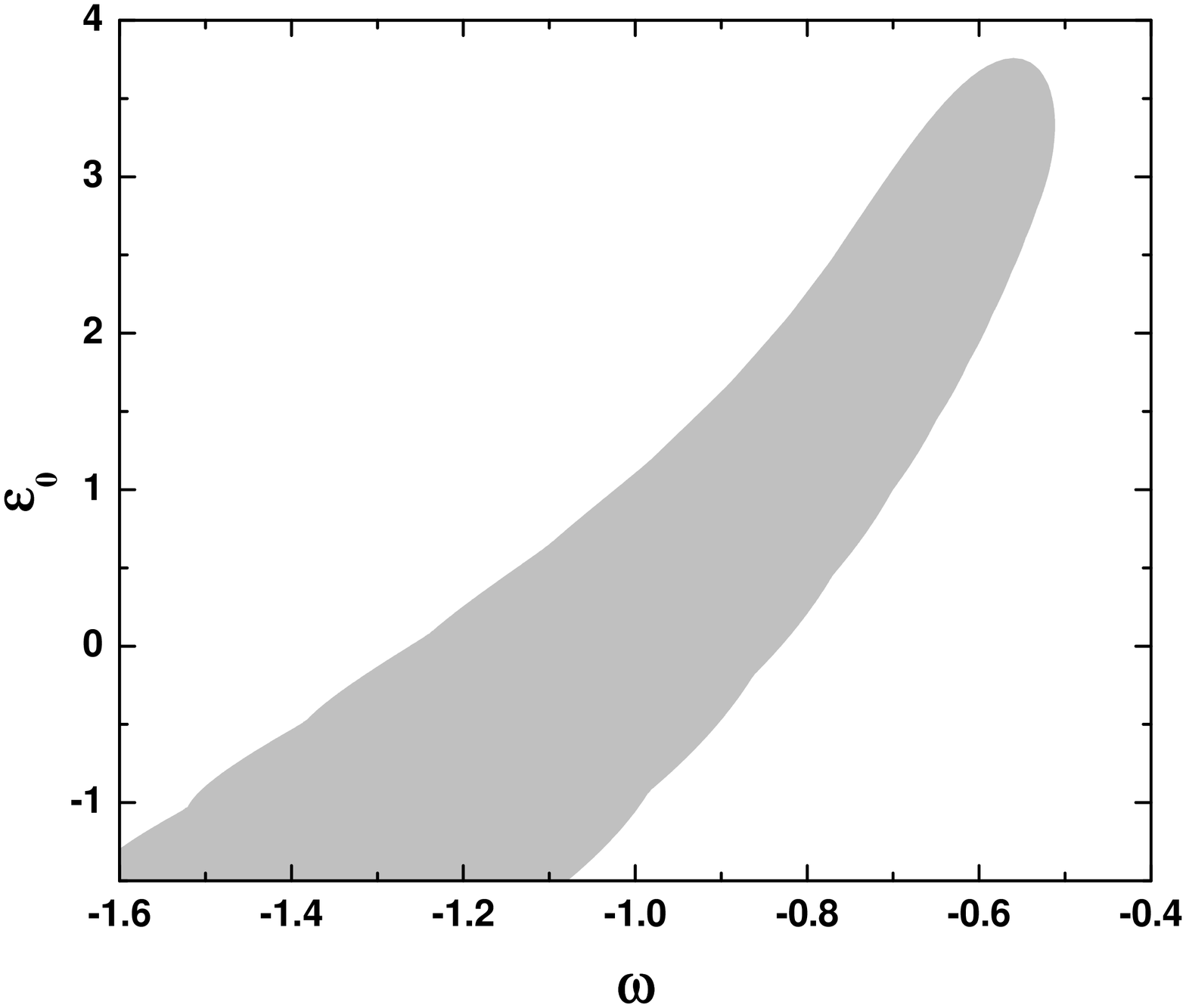,width=3.2truein,height=2.0truein,angle=-90} 
\psfig{figure=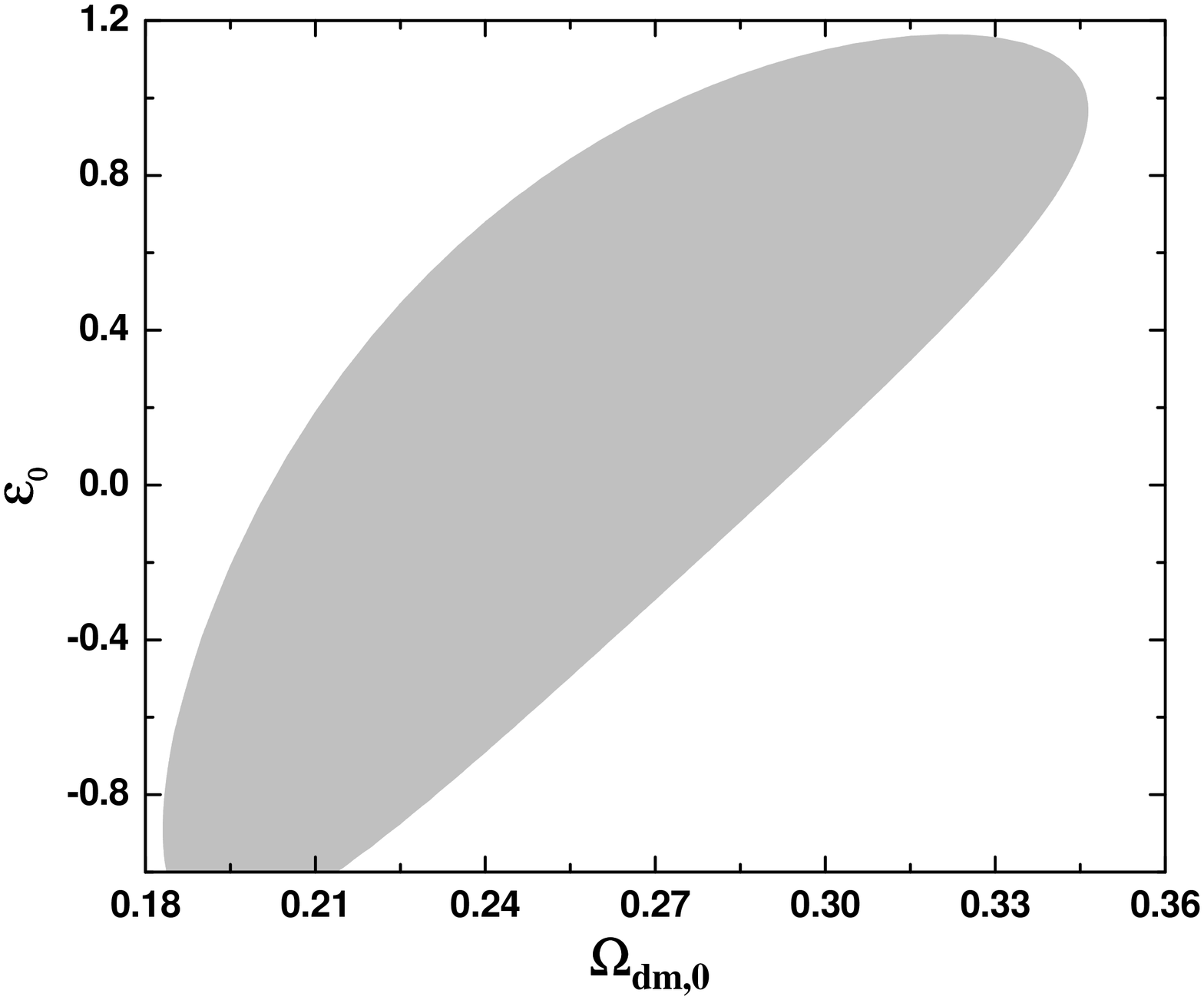,width=3.2truein,height=2.0truein,angle=-90}}
\caption{Contours of $\chi^2$ in the plane $\omega - \epsilon_{0}$ (left panel) and $\Omega_{dm,0} - \epsilon_{0}$, with $\omega = -1$, (right panel) for P2. The best fit values are $\epsilon_{0} = -1.19$ and $\omega = -1.16$ (left panel), while in the case $\Lambda(t)$CDM (right panel) we have found $\Omega_{dm,0} = 0.245$ and $\epsilon_{0}= 0.035$.}
\label{fig:qzw}
\end{figure*}

\section{Observational analysis}

As we have seen, the model here discussed comprises a multitude of cosmological solutions for different combinations of its parameters. In this section, we will discuss more quantitatively the observational aspects of this class interacting scenarios. To this end we perform a joint analysis involving current SNe Ia, CMB/BAO data. In our analysis, we fix $\Omega_{b,0} = 0.0416$ from WMAP results~\cite{cmbnew} (which is also in good agreement with the bounds on the baryonic component derived from primordial nucleosynthesis~\cite{nucleo}) and consider the recent determination of the Hubble parameter $H_0 = 74.2 \pm 4.8$~\cite{hubble} in conjunction with the CMB constraint $\Omega_{dm,0}h^2 = 0.109 \pm 0.006$~\cite{cmbnew}. We use a recent SNe Ia compilation, the so-called Union sample compiled in Ref.~\cite{union} which includes recent large samples from SNLS~\cite{snls} and ESSENCE~\cite{essence} surveys, older data sets and the recently extended data set of distant supernovae observed with the Hubble Space Telescope. The total compilation amounts to 414 SNe Ia events, which was reduced to 307 data points after selection cuts.

Following Ref.~\cite{sollerman} we use constraints derived from the product of the CMB acoustic scale 
\begin{equation}
\ell_{A} = \pi d_A (z_*)/r_s(z_*)\;,
\end{equation}
 and the measurement of the ratio of the sound horizon scale at the drag epoch to the BAO dilation scale, 
\begin{equation}
 r_s(z_d )/D_V(z_{\rm{BAO}})\;.
\end{equation}
In the previous expressions, $d_A (z_*)$ is the comoving angular-diameter distance to recombination $z_* = 1089$ and $r_s(z_*)$ is the comoving sound horizon at photon decoupling given by $r_s(z_*) = \int_{z_*}^{\infty} \frac{c_s}{H(z)} dz$, which depends upon the speed of sound before recombination $(c_s)$. $z_d \simeq 1020$ is the redshift of the drag epoch (at which the acoustic oscillations are frozen in) and the so-called dilation scale, $D_V$, is given by $D_V(z) = [czr^{2}(z)/H(z)]^{1/3}$. 

By combining the ratio $r_s (z_d = 1020)/r_s (z_*=1090) = 1.044 \pm 0.019$ ~\cite{Komatsu} with the measurements of $r_s(z_d )/D_V(z_{\rm{BAO}})$ at $z_{\rm{BAO}} = 0.20$ and 0.35 from Ref.~\cite{Percival}, Sollerman {\it et al.} (2009) found
$$
f_{0.20} = d_A (z_*)/D_V (0.2) = 17.55 \pm0.65
$$
$$
f_{0.35} = d_A (z_*)/D_V (0.35) = 10.10 \pm 0.38\;.
$$
In our analysis, we minimize the function $\chi^2_{\rm{T}} = \chi^2_{\rm{SNe}} + \chi^2_{\rm{CMB/BAO}}$, where  $\chi_{\rm{CMB/BAO}}^{2} = \left[f_{0.2}(z|\mathbf{s}) -
f_{0.2}\right]^2/\sigma_{0.2}^2 + \left[f_{0.35}(z|\mathbf{s}) - f_{0.35}\right]^2/\sigma_{0.35}^2$ and $\mathbf{s}$ stands for the model parameters. This total $\chi^2_{\rm{T}}$ function, therefore, takes into account all the observational data discussed above.

The results of our statistical analyses are shown in Figs. 4 and 5. We show $1$ and $2\sigma$ confidence regions in the parametric spaces: $\omega - \epsilon_{0}$ and $\epsilon_{0} - \xi$ for P1 and $\omega - \epsilon_{0}$ and $\Omega_{dm,0} - \epsilon_{0}$ for P2 that arise from the joint analysis described above. Note that in all panels both negative and positive values for the interacting parameter are allowed by these analyses. Physically, this amounts to saying that not only an energy flow from dark energy to dark matter ($\epsilon_{0} > 0$) is observationally allowed but also a flow from dark matter to dark energy ($\epsilon_{0} < 0$) [see Eq. (\ref{dm})]. In right panel of Fig. 4 we show the analysis for $\epsilon_{0}$ and $\xi$ by fixing the dark energy EoS at $\omega = -1$, which is fully equivalent to the vacuum decay scenario proposed in Ref.~\cite{ernandes2}. As expected, we note that the current observational bounds on $\xi$ are quite weak since it appears as a power of the scale factor in the energy density [Eqs. (\ref{dm}) and (\ref{de})]. When $\xi$ takes more negative values $\epsilon_0 \rightarrow 0$, i.e., this scenario behaves very similarly to the standard $\Lambda$CDM model. Note also that the $\Omega_{dm,0}$ parameter (Fig. 5 - right panel) is very well bounded by observations. In the case of P1, the best-fit found are $\epsilon_{0}= -0.11$ and $\omega = -1.04$ (left panel), whereas for $\Lambda(t)$CDM model (right panel) we have found $\epsilon_{0}= 0.49$ and $\xi = 0.41$. For P2 we have found the following best-fit values $\epsilon_{0} = -1.19$ and $\omega = -1.16$ (left panel), and $\Omega_{dm,0} = 0.245$ and $\epsilon_{0}= 0.035$ (right panel - $\Lambda(t)$CDM scenario).

\section{Final Remarks}

In this paper, we have investigated a general class of models with interaction in the dark sector whose evolution law of the dark energy is deduced from the effect of the same on the CDM expansion rate. Contrary to most similar analyses available in literature, we consider a more general case in which (i) the interaction term $\epsilon$ is a function of the scale factor and (ii) the EoS parameter may take any value ($w < 0$). We have shown that many previous phenomenological approaches discussed in the literature are particular cases of our approach.

We have investigated the dynamical behaviour of this scenario and found a number viable of cosmological solutions for two parametrizations of $\epsilon(a)$ (Figs. 1, 2 and 3). In the first case (P1), when $\epsilon_{0} > 0$ and the $\xi$ parameter takes large positive values ($\gtrsim 0.8$) we have found solutions with transient acceleration, in which the dark matter-dark energy interaction will drive the Universe to a new matter-dominated era in the future. P2, although depends only on the $\epsilon_0$ parameter, also supplies solutions with transient acceleration, when $\epsilon_{0} > 1.2$. As mentioned earlier, this kind of solution seems to be in agreement with theoretical constraints from String/M theories on the quintessence potential $V(\phi)$ or, equivalently, on the dark energy equation-of-state $w$, as discussed in Ref.~\cite{fischler}.

By combining  recent data of SNe Ia (Union sample) with the so-called CMB/BAO ratio at two redshifts, $z = 0.2$ and $z = 0.35$, we have shown that that both an energy flow from dark energy to dark matter as well as a flow from dark matter to dark energy are possible. For the two different parameterizations discussed here, we have also investigated $\Lambda(t)$CDM scenarios for which $\omega = -1$. We have also found that positive values of $\xi$ are largely favored over negative ones.

\begin{acknowledgments}
This work was supported by CAPES (Brazilian Research Agency). The author thanks J. S. Alcaniz for valuable discussions.
\end{acknowledgments}

\end{document}